\documentclass[10pt]{article}

\usepackage[small,bf]{caption}
\usepackage{amsfonts,amsmath,amssymb}
\usepackage{times}
\usepackage{float}
\usepackage{braket}
\usepackage{subfigure}
\usepackage{graphicx,wrapfig}

\newcommand{\e}{\ensuremath{\mathrm{e}}}
\newcommand{\GZ}{\ensuremath{\mathrm{GZ}}}
\newcommand{\M}{\ensuremath{\mathrm{M}}}

\newcommand{\p}{\partial}

\newcommand{\occ}{\overline{c}}

\renewcommand{\d}{\ensuremath{\mathrm{d}}}

\begin{document}
\title{ { \bf  The effects of Gribov copies in $2D$ gauge theories}}
\author{D.~Dudal\thanks{ddudal@mit.edu, david.dudal@ugent.be}\,\,$^{a,b}$, S.~P.~Sorella\thanks{sorella@uerj.br\;;\;Work supported by FAPERJ, Funda{\c
c}{\~a}o de Amparo {\`a} Pesquisa do Estado do Rio de Janeiro, under
the program {\it Cientista do Nosso Estado},
E-26/100.615/2007.}\,\,$^c$,
N.~Vandersickel\thanks{nele.vandersickel@ugent.be}\,\,$^b$ and
H.~Verschelde\thanks{henri.verschelde@ugent.be}\,\,$^b$\\\\
\small $^a$ Center for Theoretical Physics, Massachusetts Institute
of Technology,\\\small
77 Massachusetts Avenue, Cambridge, MA 02139, USA\\\\
\small $^b$ Ghent University, Department of Mathematical Physics and
Astronomy, \\ \small Krijgslaan 281-S9, 9000 Gent,
Belgium\\\\
$^c$ \small Departamento de F\'{\i }sica Te\'{o}rica, Instituto de
F\'{\i }sica, UERJ - Universidade do Estado do Rio de Janeiro,\\
\small   Rua S\~{a}o Francisco Xavier 524, 20550-013 Maracan\~{a},
Rio de Janeiro, Brasil \normalsize}
\date{}
\maketitle\vspace{-0.5cm}
\begin{abstract}
\noindent In previous works, we have shown that the Gribov-Zwanziger
action, which implements the restriction of the domain of
integration in the path integral to the Gribov region, generates
extra dynamical effects which influence the infrared behaviour of
the gluon and ghost propagator in $SU(N)$ Yang-Mills gauge theories.
The latter are in good agreement with the most recent lattice data
obtained at large volumes, both in $4D$ and in $3D$. More precisely,
the gluon propagator is suppressed and does not vanish at zero
momentum, while the ghost propagator keeps a $1/p^2$ behaviour for
$p^2\approx0$. Instead, in $2D$, the lattice data revealed a
vanishing zero momentum gluon propagator and an infrared enhanced
ghost, in support of the usual Gribov-Zwanziger scenario. We will
now show that the $2D$ version of the Gribov-Zwanziger action still
gives results in qualitative agreement with these lattice data, as
the peculiar infrared nature of $2D$ gauge theories precludes the
analogue of the dynamical effect otherwise present in $4D$ and $3D$.
Simultaneously, we also observe that the Gribov-Zwanziger
restriction serves as an infrared regulating mechanism.
\end{abstract} 
\vspace{-14cm} \hfill MIT-CTP 3974 \vspace{14cm}

 \noindent
\section{Introduction}
Two-dimensional, i.e. with one space and one time dimension, $SU(N)$
Yang-Mills gauge theory has been widely investigated as a kind of
toy model for real life gauge theories. E.g., in the large $N$
limit, 't Hooft has shown that confinement occurs, while mesons,
built from a quark-antiquark pair, display the analogue of ``Regge
trajectories'' \cite{'t Hooft:1974hx}.  Even if one omits the
quarks, pure $2D$ $SU(N)$ Yang-Mills gauge theory remains confining.
Although $2D$ gauge theories share some similarities with their also
confining $3D$ or $4D$ counterparts, there are nevertheless some
notable differences. Indeed, at the classical level, as the gauge
field $A_{\mu}$ contains only two degrees of freedom in $2D$,
imposing e.g. the Landau gauge condition, $\partial_{\mu} A_{\mu} =
0$, already removes these two degrees of freedom from the physical
spectrum. Therefore, as no physical degrees of freedom remain,
confinement seems to be a rather ``trivial'' phenomenon, if one sees
confinement as the absence of the elementary gluon degrees of
freedom. In contrast, in $3D$ and $4D$, one respectively two degrees
of freedom are maintained, hence confinement seems to be more than
``trivial''. Also at the quantum level, the $2D$ situation is
different from the $4D$ case. In $2D$, the coupling $g$ acquires the
dimension of a mass and thus the theory becomes highly
superrenormalizable. However, a drawback of the
superrenormalizability is the appearance  of severe infrared
instabilities and therefore an infrared regulator, usually put in by
hand, is necessary. We emphasize that caution is anyhow at place
when performing calculations in $2D$ gauge theories as discussed in
\cite{Bassetto:1999ah}. Let us also mention that certain studies
questioned some of the results of \cite{'t Hooft:1974hx} by
recalculating the fermion propagator using other infrared
regularization methods, and the corresponding results were
qualitatively
different \cite{Frishman:1976vi,Pak:1976dk,Wu:1977hi}. \\

In this letter, we shall focus on one particular aspect of $2D$
gauge theories, namely the gluon and ghost propagator, and we shall
work in the Landau gauge, as this is the most studied gauge, also
from the numerical viewpoint of lattice simulations. In particular,
in $2D$, very big lattice volumes can be achieved, so $2D$ again
serves as an interesting toy case. The propagators in the Landau
gauge have received considerable interest in $2D$, $3D$ and $4D$, as
they are expected to have a connection with confinement. Let us
enlist a few of such aspects: (1) the gluon propagator displays a
violation of positivity, signalling that transverse gluons cannot be
physical excitations. A vanishing gluon propagator at zero momentum
means a maximal positivity violation; (2) the ghost enjoys an
infrared enhancement, which according to e.g. \cite{Alkofer:2008tt}
gives rise to confinement; (3) an enhanced ghost makes the
Kugo-Ojima confinement criterion to be fulfilled
\cite{Kugo:1979gm,Kugo:1995km} (see also \cite{Braun:2007bx}).
However, in $3D$ and $4D$, recent lattice results show a ghost
propagator which does not appear to be infrared enhanced, while an
infrared positivity violating gluon propagator nonvanishing at zero
momentum is found
\cite{Cucchieri:2007md,Bogolubsky:2007ud,Cucchieri:2007rg,Cucchieri:2008fc}.
Surprisingly, in $2D$, the ghost propagator still displays an
enhanced behavior while the gluon propagator does vanish at the
origin \cite{Cucchieri:2007rg,Cucchieri:2008fc,Maas:2007uv}.\\

In recent work \cite{Dudal:2007cw,Dudal:2008sp}, we have
exhaustively examined the $4D$ case within the extended
Gribov-Zwanziger framework, that relies on the original
Gribov-Zwanziger action enlarged with an extra mass term while
preserving its locality and renormalizability. This mass was fixed
in a variational way, and as such represented an additional
nontrivial dynamical effect. For the benefit of the reader, let us
first briefly summarize this framework. We recall that the Landau
gauge condition, $\p_\mu A_\mu=0$, does not uniquely fix the local
gauge freedom, there are still gauge equivalent fields
$\widetilde{A}_\mu$ which are also transverse, $\p_\mu
\widetilde{A}_\mu=0$ \cite{Gribov:1977wm}. As a consequence, the
domain of integration in the path integral has to be restricted in a
suitable way. Gribov proposed to restrict the domain of integration
to the Gribov region $\Omega$. Within this region $\Omega$, the
Faddeev-Popov operator ${\cal M}^{ab}\equiv-\partial _{\mu }\left(
\partial _{\mu }\delta ^{ab}+gf^{acb}A_{\mu }^{c}\right) $ is positive definite, i.e.
${\cal M}^{ab} >0$, while at the boundary $\partial \Omega$ of this
region, the first Gribov horizon, the first vanishing eigenvalue of
${\cal M}^{ab}$ appears \cite{Gribov:1977wm}. In this fashion, a
large set of gauge copies is excluded, as their existence is related
to the presence of zero modes\footnote{Parametrizing a gauge
transformation with an infinitesimal gauge parameter $\omega^{a}$, a
gauge equivalent field $\widetilde{A}_\mu$ is given by
$\widetilde{A}_\mu^a=A_\mu^a-D_\mu^{ab}\omega^b$. Hence, $\p_\mu
\widetilde{A}_\mu=\p_\mu A_\mu=0$ leads to $\p_\mu
D_\mu^{ab}[A]\omega^b=0$, i.e. $\omega^a$ represents a zero mode of
${\cal M}^{ab}$.} of ${\cal M}^{ab}$. Gribov implemented his idea at
the semi-classical level \cite{Gribov:1977wm}, and later Zwanziger
has been able to implement the restriction to $\Omega$ at all orders
through the introduction of a nonlocal horizon function appearing in
the Boltzmann weight defining the Euclidean Yang-Mills measure
\cite{Zwanziger:1989mf,Zwanziger:1992qr}. It is worth remarking that
the Gribov region itself is also not free from gauge copies
\cite{Semenov,Dell'Antonio:1991xt,Dell'Antonio2,vanBaal:1991zw}. To
avoid these extra copies, a further restriction to an even smaller
region $\Lambda$, known as the fundamental modular region, should be
implemented. Unfortunately, it is unknown how this goal can be
achieved. It is not unexpected that a restriction to the Gribov
region $\Omega$, and thus on the allowed gauge field configurations,
has a strong influence on the behaviour of the propagators in the
infrared, as found for the first time in \cite{Gribov:1977wm}: the
ghost propagator gets enhanced in the infrared, while the gluon
propagator is suppressed and goes to zero at zero momentum. As
already mentioned, this does not seem to be supported anymore by the
most recent lattice data. We recently introduced a refined version
of the Gribov-Zwanziger framework and consequently found a ghost
propagator which was no longer enhanced and a gluon propagator which
was nonvanishing at zero momentum, both in accordance with the
latest $4D$ lattice data \cite{Dudal:2007cw,Dudal:2008sp}. Also in
$3D$, similar results were found \cite{Dudal:2008rm}. Naturally, the
question rises whether a distinct result would be found in $2D$,
still within this extended Gribov-Zwanziger
framework?\\

The purpose of this letter is to present the answer to that last
query. The gluon and the ghost propagator are investigated in detail
and we shall demonstrate why the $2D$ case varies from the $3D$ and
$4D$ case from the Gribov-Zwanziger viewpoint. The paper is
organized as follows. In section \ref{sectie2}, we provide a short
overview of the ordinary Gribov-Zwanziger action in two dimensions,
as well as of the refined Gribov-Zwanziger action,  obtained through
the inclusion of an extra mass term. In Section 3 we present two
arguments of why this new mass term, which can be consistently
introduced in $3D$ and $4D$, induces infrared instabilities in $2D$
which prevent its introduction. Firstly, we shall see that the value
of a certain condensate is already infinite at the perturbative
level when the new mass term is present. Secondly, we will also
explicitly show that the ghost self energy develops an infrared
singularity in the presence of the new mass, which (1) invalidates
any finite order approximation and more importantly, (2) enforces
one to cross the Gribov horizon $\p\Omega$, thus to leave the Gribov
region $\Omega$, which was the starting point of the whole
Gribov-Zwanziger construction. Both phenomena are related to the
infrared peculiarities of $2D$ gauge theories. Therefore, the
introduction of the novel mass term in $2D$ turns out to be
jeopardized by these infrared instabilities. As a consequence, the
ghost propagator will keep displaying an enhanced behavior and the
gluon propagator will vanish at zero momentum, in agreement with the
lattice results. Schwinger-Dyson results consistent with this $2D$
scenario can be found in
\cite{Zwanziger:2001kw,Maas:2004se,Huber:2007kc}. Let us also
mention that the usual restriction to the Gribov region regularizes
the theory in a natural way in the infrared at least at one loop
level. We end this paper with a discussion in section 4.

\section{Survey of the (extended) Gribov-Zwanziger action \label{sectie2}}
\subsection{The ordinary Gribov-Zwanziger action}
We shall start this section with a short overview of the ordinary
Euclidean Gribov-Zwanziger action in two dimensions in the Landau
gauge, and of its extended version which we originally proposed in
\cite{Dudal:2007cw}. We shall not go into any details, as it is
quite analogous to the $3D$ or $4D$ situation.

In its original nonlocal formulation, the Gribov-Zwanziger action is
given by
\begin{eqnarray}\label{actienonlocal}
S_{\mathrm{h}}&=& S_{\mathrm{YM}} + S_{\mathrm{gf}} + S_{\gamma} \;,
\end{eqnarray}
with $S_{\mathrm{YM}}$ the classical Yang-Mills action,
\begin{eqnarray}
S_{\mathrm{YM}} &=& \frac{1}{4}\int \d^2 x F_{\mu\nu}^a F_{\mu\nu}^a
\;,
\end{eqnarray}
and $S_{\mathrm{gf}}$ the gauge fixing and ghost part,
\begin{eqnarray}
S_{\mathrm{gf}} &=& \int \d^{2}x\;\left( b^{a}\partial_\mu
A_\mu^{a}+\overline{c}^{a}\partial _{\mu } D_{\mu }^{ab}c^b \right)
\end{eqnarray}
which implements the Landau gauge condition, $\p_\mu A_\mu^a=0$.
Furthermore, $S_{\gamma}$ contains the horizon function $h(x)$,
\begin{eqnarray}
S_{\gamma}\ =\ \gamma^4 \int \d^2 x\ h(x) &=&  \gamma^4  \int \d^2 x
\left( g^2 f^{abc} A^b_{\mu} \left(\mathcal{M}^{-1}\right)^{ad}
f^{dec} A^e_{\mu} \right) \;.
\end{eqnarray}
The so-called Gribov (mass) parameter $\gamma$ is determined by the
horizon condition,
\begin{eqnarray}\label{horizoncondition}
\braket{h(x)} &=& d (N^2 -1) \;,
\end{eqnarray}
with $d$ the number of space time dimensions. This action
$S_{\mathrm{h}}$ with the horizon condition \eqref{horizoncondition}
implemented, automatically restricts the gauge field configurations
to the Gribov region $\Omega$. We refer to
\cite{Zwanziger:1989mf,Zwanziger:1992qr} for more details on this
matter. As a nonlocal action is hard to be handled in a consistent
way, it would be advantageous if $S_{\mathrm{h}}$ could be
reformulated into an equivalent local version. Luckily, this goal
can be achieved by introducing a suitable set of additional fields,
leading to \cite{Zwanziger:1992qr}
\begin{eqnarray}
S_{\GZ} &=&S_{0} +  S_{\gamma}\;, \label{actielocal}
\end{eqnarray}
with
\begin{eqnarray}\label{snul}
S_{0} &=&S_{\mathrm{YM}}+ S_{\mathrm{gf}} +\int \d^{2}x\left(
\overline{\varphi }_{\mu}^{ac}\partial _{\nu}D_{\nu }^{ab}\varphi
_{\mu} ^{ac}-\overline{\omega }_{\mu}^{ac}\partial _{\nu}D_{\nu
}^{ab}\omega_{\mu} ^{ac} -g\left( \partial _{\nu }\overline{\omega
}_{\mu}^{ac}\right) f^{abm}\left( D_{\nu
}c\right)^{b}\varphi _{\mu}^{mc} \right) \;, \nonumber\\
S_{\gamma}&=& -\gamma ^{2}g\int\d^{2}x\left( f^{abc}A_{\mu
}^{a}\varphi _{\mu }^{bc}+f^{abc}A_{\mu}^{a}\overline{\varphi }_{\mu
}^{bc} + \frac{2}{g}\left(N^{2}-1\right) \gamma^{2} \right) \;,
\end{eqnarray}
where $\left( \overline{\varphi
}_{\mu}^{ac},\varphi_{\mu}^{ac}\right) $ and $\left(
\overline{\omega }_{\mu}^{ac},\omega_{\mu}^{ac}\right) $ are a pair
of complex conjugate bosonic, respectively anticommuting, fields. In
this local framework, the horizon condition \eqref{horizoncondition}
is converted to
\begin{eqnarray}\label{gapgamma}
\frac{\p \Gamma}{\p \gamma^2} &=& 0 \;,
\end{eqnarray}
with $\Gamma$ the quantum effective action,
\begin{eqnarray}
\e^{-\Gamma} &=& \int \d\Phi \e^{-S} \;.
\end{eqnarray}
Before closing this subsection, we mention that the fields, except
for $b^a$, are dimensionless while, in two dimensions, the coupling
$g$ has the dimension of a mass. Consequently, the theory is
ultraviolet superrenormalizable. On the other hand, in the infrared
region, serious problems can occur. Indeed, in perturbation theory,
higher powers of $g^2$ shall induce increasing powers of momentum in
the denominator, which will give rise to severe problems upon
integration around zero momentum. We shall come back to this issue
in section 3.

\subsection{The extended Gribov-Zwanziger action}
By analogy with previous works in four and three dimensions
\cite{Dudal:2007cw,Dudal:2008sp,Dudal:2008rm}, we shall add a mass
term of the form $M^2 \int d^2 x\;\left( \overline{\varphi}_\mu^{ab}
\varphi_\mu^{ab} -\overline{\omega}_\mu^{ab} \omega_\mu^{ab}
\right)$ to the localized Gribov-Zwanziger action $S_\GZ$. Only
later on this paper, we shall demonstrate that including this mass
term will give rise to infrared instabilities. However, purely from
the algebraic and dimensional viewpoint, this mass term cannot be
excluded in $2D$ just as in $3D$ or $4D$
\cite{Dudal:2007cw,Dudal:2008sp}. We recall that in the three and
four dimensional case, this mass term was initially added to alter
the gluon propagator, which can be intuitively understood. Indeed,
already at the quadratic level of the action $S_\GZ$, one observes
an $A\varphi$-coupling. Therefore, changing the dynamics of the
$\varphi$-sector by adding an extra term, will affect the gluon
sector. Also the ghost propagator was modified by the addition of
this novel mass term \cite{Dudal:2007cw,Dudal:2008sp}.

Completely analogous as in 3 or 4 dimensions, one can formally prove
the (ultraviolet) renormalizability of the action making use of the
algebraic renormalization formalism and of the many Ward identities
constraining the quantum version of the action \cite{Piguet:1995er}.
We refer to our previous work \cite{Dudal:2008sp} for all the
necessary details. Of course, since there are no ultraviolet
infinities, renormalization is in principle trivial. However, the
algebraic formalism allows us to discuss more than just the form of
the (potential) counterterm. For example, we also used it in
\cite{Dudal:2008sp} to study the Slavnov-Taylor identities in the
presence of the restriction to the Gribov region $\Omega$. We recall
that we have proven in \cite{Dudal:2008sp} that this restriction
necessarily spoils the BRST symmetry, but nevertheless one can still
write down a powerful set of Slavnov-Taylor identities, which
enabled us to prove the ultraviolet renormalizability in $3D$ or
$4D$.

\section{Two reasons why the refined Gribov-Zwanziger action is excluded in $2D$}
In this section, we shall provide two reasons why it is not possible
to add the novel mass $\propto
\overline{\varphi}\varphi-\overline{\omega}\omega$ to the standard
Gribov-Zwanziger action \eqref{actielocal}. It shall become clear
that it is exactly the fact that we are working in $2D$ which does
signal us that the theory with $
\overline{\varphi}\varphi-\overline{\omega}\omega$ coupled to it is
not well defined.

To start with, let us write down again the complete refined
Gribov-Zwanziger action,
\begin{eqnarray}
S' &=&S_{\GZ}+S_\M\;, \nonumber\\
S_\M&=&-M^2\int
\d^dx\left(\overline{\varphi}_\mu^{ac}\varphi_\mu^{ac}-\overline{\omega}_\mu^{ac}\omega_\mu^{ac}\right)+\int
\d^dx\left(d\frac{N^2-1}{g^2N}\varsigma M^2\lambda^2\right)\;.
\label{actielocal2}
\end{eqnarray}
The role of vacuum term proportional to the dimensionless parameter
$\varsigma$ is a bit redundant in the $2D$ case, as the problems we
shall encounter are neither related to nor curable by this quantity
$\varsigma$, which played a pivotal role in $3D$ or $4D$
\cite{Dudal:2008sp}. For completeness and comparability with the
$3D$ or $4D$ case, we have included it nevertheless.

Let us also give here our notational conventions for the gluon
propagator,
\begin{equation}\label{glpr}
    \braket{A_\mu^a A_\nu^b}_p=\mathcal{D}(p^2)\left(\delta_{\mu\nu}-\frac{p_\mu
    p_\nu}{p^2}\right)\delta^{ab}\;,
\end{equation}
and ghost propagator,
\begin{equation}\label{glpr}
    \braket{c^a \occ^b}_p=\mathcal{G}(p^2)\delta^{ab}\:,
\end{equation}
in momentum space in the Landau gauge.

Subsequently, we compute the one loop quantum effective action
$\Gamma$ as
\begin{eqnarray}
\hspace{-0.4cm}\Gamma &=&
-d(N^{2}-1)\frac{\lambda^{4}}{2g^2N}+\frac{(N^{2}-1)}{2}\left(
d-1\right) \int \frac{\d^{d}p}{\left(2\pi \right) ^{d}}\ln  \left[
p^2 \left( p^{2} + \frac{\lambda^4}{p^2+
M^2}\right)\right]+d\frac{N^2-1}{g^2N}\varsigma M^2\lambda^2\;,
\end{eqnarray}
hence the gap equation \eqref{gapgamma} is determined by
\begin{eqnarray}\label{gappie0}
\frac{2}{g^2N}&=&\int\frac{\d^2p}{(2\pi)^2}\frac{1}{p^4+M^2p^2+\lambda^4}+\frac{2}{g^2N}\varsigma
\frac{M^2}{\lambda^2}
\end{eqnarray}
for $d=2$.

\subsection{The first reason why $\overline{\varphi}\varphi-\overline{\omega}\omega$ is problematic in $2D$}
Let us recall why we originally started the study of the dynamical
effects associated to the operator
$\overline{\varphi}\varphi-\overline{\omega}\omega$ in $3D$ and $4D$
\cite{Dudal:2007cw,Dudal:2008sp}. Since the restriction to the
Gribov region introduces a massive parameter $\gamma^2$ into the
theory, it might be natural to expect a nonvanishing vacuum
expectation value for the operator
$\overline{\varphi}\varphi-\overline{\omega}\omega$ already at the
perturbative level, namely
$\braket{\overline{\varphi}\varphi-\overline{\omega}\omega}\propto\gamma^2$.
This was confirmed by explicit calculations in \cite{Dudal:2008sp}.
We then used a variational approach, expressed through the mass
$M^2$ coupled to the action, in order to take into account the
potential effects related to this operator on e.g. the gluon and
ghost propagator.

We shall now verify that our original rationale behind the study of
$\overline{\varphi}\varphi-\overline{\omega}\omega$ no longer
applies in $2D$, showing that this operator cannot be consistently
introduced in $2D$. It should not come as a too big surprise that
the difficulties related to the operator
$\overline{\varphi}\varphi-\overline{\omega}\omega$ rely on the
appearance of infrared instabilities, typical of $2D$, which
prevents the analogue phenomenon as in $3D$ or $4D$ to occur in
$2D$.

Let us take a look at the condensate
$\langle\overline{\varphi}\varphi-\overline{\omega}\omega
\rangle$. We define the energy functional as
\begin{equation}\label{W}
  \e^{-W(J,\gamma^2)}=\int \d\Psi \e^{-S_{\GZ}+\int \d^2x
    J(\overline{\varphi}\varphi-\overline{\omega}\omega)+\varsigma'J\lambda^2}\;.
\end{equation}
Here, we suitably rescaled $\varsigma$ into $\varsigma'$ for
notational convenience, $\varsigma' = d \frac{N^2 -1}{g^2 N}
\varsigma $. We have also replaced the mass $M^2$ by the more
conventional notation for a source, i.e.~$J$.

Nextly, let us consider the perturbative value of the condensate,
which is explicitly given by
\begin{equation}\label{pot4}
    \braket{\overline{\varphi}\varphi-\overline{\omega}\omega}_{\mathrm{pert}}=-\left.\frac{\p W}{\p
    J}\right\vert_{J=0}-\varsigma'\lambda^2\,.
\end{equation}
To calculate this quantity we evaluate the one loop energy
functional,
\begin{eqnarray}\label{wj}
W(J) &=& -d(N^{2}-1)\gamma^{4}+\frac{(N^{2}-1)}{2}\left( d-1\right)
\int \frac{\d^{d}p}{\left(2\pi \right) ^{d}}\ln  \left[ p^2 \left(
p^{2} + \frac{\lambda^4}{p^2+ J}\right)\right]-\varsigma'\lambda^2
\;.
\end{eqnarray}
With the help of dimensional regularization we find the following
finite result,
\begin{eqnarray}
W(J) &=&  - \frac{\lambda^4}{g^2 N}(N^2 -1) -\frac{N^2 - 1}{16 \pi}
\left[ J \ln \frac{4 \lambda^4}{J^2} - \sqrt{J^2 - 4
\lambda^4}\ln\frac{J- \sqrt{J^2 - 4 \lambda^4}}{J+\sqrt{J^2 - 4
\lambda^4}} \right]-\varsigma'\lambda^2\,.
\end{eqnarray}
This expression is well-defined when taking the limit $J\to0$. This
corresponds to the pure Gribov-Zwanziger case, where $M^2 = J =0$.
However, the derivative w.r.t.~$J$ is singular for $J = 0$. Indeed,
we find
\begin{eqnarray}
\frac{ \partial W(J)}{ \partial J}  &=&  -\frac{N^2 - 1}{16 \pi}
\left[ \frac{-J}{\sqrt{J^2 - 4 \lambda^4}} \ln\frac{J- \sqrt{J^2 - 4
\lambda^4}}{J+\sqrt{J^2 - 4 \lambda^4}}  + \ln \frac{4
\lambda^4}{J^2} \right]-\varsigma'\lambda^2\;,
\end{eqnarray}
in which the second term diverges for $J\to0$. This would imply that
\begin{eqnarray}
 \Braket{\overline{\varphi}\varphi-\overline{\omega}\omega }= \infty\;.
\end{eqnarray}
This strongly suggests that is it impossible to couple the operator
to the theory without even causing pathologies already in
perturbation theory. A way to appreciate that this divergence is
stemming from the infrared region is to derive first expression
\eqref{wj} w.r.t. $J$ (assuming this is allowed) and then set $J=0$,
in which case
\begin{eqnarray}
\left.\frac{ \partial W(J)}{ \partial J}\right\vert_{J=0}
&=&\frac{N^2-1}{2}(d-1)\left(\int\frac{\d^dp}{(2\pi)^d}\frac{p^2}{p^4+\lambda^4}-\int\frac{\d^dp}{(2\pi)^d}\frac{1}{p^2}\right)-\varsigma'\lambda^2\;.
\end{eqnarray}
The second term in the previous expression is typically zero in
dimensional regularization, \emph{except} when $d=2$ as it then
develops an infrared pole.

Having revealed a first counterargument against the introduction of
the mass operator
$M^2(\overline{\varphi}\varphi-\overline{\omega}\omega)$ in $2D$,
let us give an even stronger objection in the following subsection.

\subsection{The second (main) reason why $\overline{\varphi}\varphi-\overline{\omega}\omega$ is problematic in $2D$: the ghost propagator}
\subsubsection*{The case $M^2 \not= 0$}
Let us consider the one loop ghost propagator displayed in Figure 1, which yields
\begin{eqnarray}\label{gh}
{\cal G}^{ab}(k) &=& \delta^{ab} \frac{1}{k^2}\frac{1}{1-\sigma(k)} \;,
\end{eqnarray}
after resummation into the one loop ghost self energy.
\begin{figure}[t]\label{ghost}
    \begin{center}
        \scalebox{1}{\includegraphics{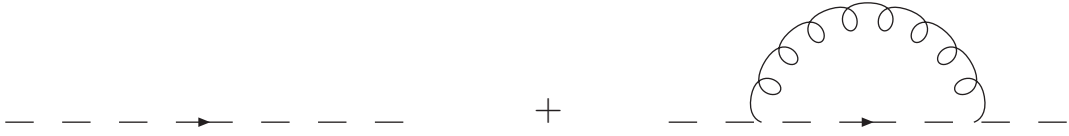}}
        \caption{The one loop ghost propagator.}
    \end{center}
\end{figure}
Explicitly, the one loop correction to the ghost self energy reads
\begin{eqnarray}\label{sigmadef}
\sigma(k) &=& g^2N \frac{k_{\mu} k_{\nu}}{k^2} \int \frac{\d^2
q}{(2\pi)^2} \frac{1}{(k-q)^2} \frac{q^2 + M^2}{q^4 + M^2q^2 +
\lambda^4}\left(\delta_{\mu\nu}-\frac{q_\mu q_\nu}{q^2}\right)\;.
\end{eqnarray}
We recall here that the ghost self energy correction $\sigma(k)$ can
be used as a kind of ``order parameter'' to check whether a gauge
configuration lies inside or outside the Gribov horizon. Indeed, the
ghost propagator is positive definite inside $\Omega$ by
construction, meaning that $\sigma(k)\leq 1$. As a matter of fact,
the requirement that $\sigma(k)\leq 1$ is usually called the no-pole
condition, and it played a key role in Gribov's original
implementation of the restriction to the region $\Omega$
\cite{Gribov:1977wm,Sobreiro:2005ec}.

Looking at the integral \eqref{sigmadef}, the term
$\sim\frac{1}{(q-k)^2}$ which could potentially lead to an infrared
singularity upon integration, is partially ``protected'' by the
external momentum $k$. One might expect that the infrared divergence
will only reveal itself in the limit $k\to0$.

Bearing this in mind, let us determine $\sigma(k)_{k^2\sim0}$ by
performing the $\vec{q}$-integration in \eqref{sigmadef} exactly for
an arbitrary momentum $\vec{k}$. We shall invoke polar coordinates.
Without loss of generality, we can put the $q_x$-axis along
$\vec{k}$ to write
\begin{eqnarray}
\sigma(k) &=& \frac{g^2N}{4\pi^2}\int_0^{\infty}q\d
q\frac{q^2+M^2}{q^4+M^2q^2+\lambda^4}\int_{0}^{2\pi}\d\theta
\frac{1}{k^2+q^2-2qk\cos\theta}(1-\cos^2\theta)\;,
\end{eqnarray}
where we made use of $\vec{k}\cdot\vec{q}=kq\cos\theta$. The
Poisson-like $\theta$-integral can be easily calculated using a
contour integration,
\begin{equation}
    \int_{0}^{2\pi}\d\theta
\frac{1-\cos^2\theta}{k^2+q^2-2qk\cos\theta}=\left\{\begin{array}{c}
                                                    \frac{\pi}{q^ 2}\qquad\mbox{if\;} k^2\leq q^2 \\
                                                    \frac{\pi}{k^ 2}\qquad\mbox{if\;} q^2\leq k^2
                                                  \end{array}\right.\;,
\end{equation}
so we obtain
\begin{eqnarray}\label{sigma2}
\sigma(k) &=&
\frac{g^2N}{4\pi}\left(\frac{1}{k^2}\int_0^{k}\frac{q(q^2+M^2)}{q^4+M^2q^2+\lambda^4}\d
q+\int_k^{\infty}\frac{q^2+M^2}{q(q^4+M^2q^2+\lambda^4)}\d
q\right)\;.
\end{eqnarray}
It appears that both integrals are well-behaved in the infrared and
ultraviolet for $k>0$.

Notice that we did not invoke the gap equation \eqref{gappie0} yet.
This is possible, but neither necessary nor instructive at this
point. In order to have a better understanding of the $k\to0$
behaviour, we can calculate the integrals in \eqref{sigma2}, and
extract the small momentum behaviour. Doing so, one finds
\begin{equation}\label{sigma3}
    \left.\sigma(k)\right|_{k^2\sim 0}\sim
    -\frac{g^2N}{8\pi}\frac{M^2}{\lambda^4}\ln(k^2)
\end{equation}
in the case that $M^2\neq0$, which is a well-defined result, in
contrast with \eqref{sigmang}.

However, there is still an infrared instability in the theory due to
the final $\ln(k^2)$-factor appearing in $\sigma(k)$ for small $k$.
This is our second main argument why coupling the mass operator
$(\overline{\varphi}\varphi-\overline{\omega}\omega)$ to the theory
causes problems:
\begin{itemize}
\item The quantum correction to the self energy explodes for small
$k$, completely invalidating the loop expansion. This problem does
not occur in $3D$ or $4D$, since there $\sigma\leq 1$. It is not
difficult to imagine that the infrared $\ln(k^2)$-singularity will
spread itself through the theory, making everything ill-defined for
small $k$.
\item Moreover, we also encounter a problem of a more
fundamental nature. The starting point of the whole
construction was to always stay within the Gribov
horizon $\Omega$. This can be assured by the so
called no-pole condition, i.e.~$\sigma(k^2) \leq 1$
as stated in the original article by
Gribov \cite{Gribov:1977wm}. Since $M^2$ must be
positive\footnote{A negative $M^2$ would lead to tachyonic
instabilities in the theory, see e.g. the vacuum functional as an
example.}, we clearly see from \eqref{sigma3} that
\begin{equation}
\left.\sigma(k)\right|_{k^2\sim0}\gg1\;,
\end{equation}
hence \eqref{gh} is signalling us that we have crossed the horizon.
\end{itemize}
This confirms again that $M^2=0$ is the only viable option, i.e. we
cannot go beyond the standard Gribov-Zwanziger action if we want to
avoid the appearance of destructive infrared issues, which
unavoidably force the theory to leave the Gribov region.

\textbf{Remark.} In the previous paragraph, in order to calculate
\eqref{sigmadef}, we have first determined the integral in
expression \eqref{sigmadef} exactly and then we have taken the limit
$k^2 \to 0$. However, one usually
\cite{Gribov:1977wm,Sobreiro:2005ec} first expands the integrand for
small $k^2$ and then performs the loop integration, as this
considerably reduces the calculational effort.  In the current case,
this course of action unfortunately leads to incorrect results.
Indeed, doing so, we would reexpress ``1'' as
\begin{eqnarray}
  1 &=& g^2N\frac{k_\mu k_\nu}{k^2}\int \frac{\d^2q}{(2\pi)^2} \frac{1}{q^4+M^2q^2+\lambda^4}\left(\delta_{\mu\nu}-\frac{k_\mu k_\nu}{k^2}\right)+\varsigma\frac{M^2}{\lambda^2}\;,
\end{eqnarray}
an operation which is based on the gap equation \eqref{gappie0}.
Subsequently we rewrite $1-\sigma(k)$,
\begin{eqnarray}
  1-\sigma(k) &=& g^2N\frac{k_\mu k_\nu}{k^2}\int \frac{\d^2q}{(2\pi)^2}
  \frac{1}{q^4+\lambda^4}\left(1-\frac{q^2}{(k-q)^2}\right)\left(\delta_{\mu\nu}-\frac{k_\mu
  k_\nu}{k^2}\right)+\varsigma \frac{M^2}{\lambda^2}\nonumber\\
&+&g^2N \frac{k_{\mu} k_{\nu}}{k^2} \int \frac{\d^2 q}{(2\pi)^2}
\frac{1}{(k-q)^2} \frac{M^2}{q^4 + M^2q^2 +
\lambda^4}\left(\delta_{\mu\nu}-\frac{q_\mu q_\nu}{q^2}\right)
  \;,
\end{eqnarray}
and then we expand the integrand\footnote{We notice that there will
be no terms of odd order in $k$, since this would correspond to an
odd power of $q$, which will vanish upon integration due to
reflection symmetry.} around $k^2\sim0$ to find at lowest order,
\begin{eqnarray}\label{sigmang}
  \left.(1-\sigma(k))\right\vert_{k^2\sim0} &=&\frac{g^2N}{2}\int \frac{\d^2q}{(2\pi)^2}\frac{M^2}{q^2(q^4+M^2q^2+\lambda^4)} +\varsigma\frac{M^2}{\lambda^2}+\mathcal{O}(k^2)\;.
\end{eqnarray}
From this expression, we are led to believe that $1-\sigma(k)$,
hence $\sigma(k)$, is ill-defined at small $k^2$, due to an infrared
singularity which makes the integral in the r.h.s. of
\eqref{sigmang} to explode. However, this is not true, as in this
case, the limit and the integration cannot be exchanged. The only
correct way is to first calculate the integral and then take the
limit as was done in the previous paragraph. Further on this
section, we shall explicitly explain why expression \eqref{sigmang}
is wrong by exploring the $M^2 =0$ case in more detail.

\subsubsection*{The case $M^2 =0$}
It is instructive to take a closer look at the usual
Gribov-Zwanziger scenario.  One finds for $M^2=0$ that
\begin{equation}\label{sigma3}
    \left.\sigma(k)\right|_{k^2\sim 0}\sim
    \frac{g^2N}{4\pi}\left(\frac{\pi}{4\lambda^2}-\frac{k^2}{4\lambda^4}\right)\;,
\end{equation}
a result which is indeed free of infrared instabilities. We also
point out that ordinary (perturbative) Yang-Mills theory is
recovered when $\lambda=0$. It is hence nice to observe that this
again causes troubles in the infrared since the $\lambda\to0$ limit
diverges. This is just a manifestation of the fact that $2D$ gauge
theories are infrared sick at the perturbative level, and need some
(dynamical) regularization. Apparently, at least at the level of the
ghost propagator at one loop, the Gribov mass acts a natural
regulator in the infrared sector.

We should still use the gap equation in \eqref{sigma3} to find the
correct ghost propagator. The gap equation \eqref{gapgamma} for
$M^2=0$ is readily computed as
\begin{eqnarray}\label{gappie}
\frac{2}{g^2 N} &=& \int \frac{\d^2 p}{(2\pi)^2} \frac{1}{p^4 +
\lambda^4} = \frac{1}{8 \lambda^2}\;.
\end{eqnarray}
Evoking this gap equation, we find
\begin{eqnarray}\label{zinvol}
1-\sigma(k) &=& 1
-\frac{g^2N}{4\pi}\left(\frac{\pi}{4\lambda^2}-\frac{k^2}{4\lambda^4}\right)=
\frac{g^2N}{4\pi}\frac{k^2}{4\lambda^4}=\frac{1}{\pi g^2N}\;.
\end{eqnarray}
Henceforth, we obtain
\begin{eqnarray}\label{gh2}
\left.{\cal G}^{ab}(k)\right|_{k^2\sim0}  &=& \left.\delta^{ab}
\frac{1}{k^2} \frac{1}{1-\sigma(k)}\right|_{k^2\sim0}=\frac{\pi
g^2N}{k^4}\;.
\end{eqnarray}
We conclude that the ghost propagator is clearly enhanced and displays the typical behavior $\sim 1/k^4$ in the deep infrared, in accordance with the usual Gribov-Zwanziger scenario.\\

\textbf{Remark.} As we already announced earlier in this section,
let us have a closer look at the $M^2=0$ case. In a way completely
similar to the $M^2\neq0$ case, we find, around $k^2\sim0$,
\begin{eqnarray}
  \left.(1-\sigma(k))\right\vert_{k^2\sim0} &=& g^2N\frac{k_\mu k_\nu}{k^2}\int \frac{\d^2q}{(2\pi)^2}
  \frac{1}{q^4+\lambda^4}\left(\frac{k^2}{q^2}-4\frac{(k\cdot q)^2}{q^2}\right)\left(\delta_{\mu\nu}-\frac{k_\mu
  k_\nu}{k^2}\right)+\mathcal{O}(k^4)\;,
\end{eqnarray}
where we have expanded the integrand w.r.t. $q$ before integrating.
Exploiting polar coordinates once more, we are now brought to
\begin{eqnarray}
  \left.(1-\sigma(k))\right\vert_{k^2\sim0} &=& \frac{g^2N}{4\pi^2}k^2\int_{0}^{+\infty}\frac{q\d
  q}{q^2}\frac{1}{q^4+\lambda^4}\int_0^{2\pi}(1-4\cos^2\theta)(1-\cos^2\theta)\d\theta+\mathcal{O}(k^4)\;.
\end{eqnarray}
Surprisingly, the $\theta$-integral vanishes, as it can be easily
checked. In fact, one can extend this observation to all orders in
$k$. To do so, we write
\begin{eqnarray}
\frac{q^2}{(q-k)^2}=\frac{q^2}{q^2+k^2-2qk\cos\theta}=\frac{1}{1+\frac{k^2}{q^2}-2\frac{k}{q}\cos\theta}=\sum_{n=0}^{\infty}\left(\frac{k}{q}\right)^n{\cal
U}_n(\cos\theta)\;,
\end{eqnarray}
where we introduced the Chebyshev polynomials of the second kind,
${\cal U}_n(x)$. It holds that \cite{wolfram}
\begin{eqnarray}\label{cheb}
  {\cal U}_n(\cos\theta) &=& \frac{\sin((n+1)\theta)}{\sin\theta}\;.
\end{eqnarray}
Subsequently, we can rewrite
\begin{eqnarray}
1-\sigma(k)&=& g^2N \int
\frac{\d^2q}{(2\pi)^2}\sum_{n=1}^{\infty}(1-\cos^2\theta){\cal
U}_n(\cos\theta)\left(\frac{k}{q}\right)^n\frac{1}{q^4+\lambda^4}\;,
\end{eqnarray}
where use has been made of ${\cal U}_0(x)=1$. Assuming that the
integral and the infinite sum can be interchanged, we are led to
\begin{eqnarray}\label{slecht}
1-\sigma(k)&=& \frac{g^2N}{4\pi^2}
\sum_{n=1}^{\infty}k^n\int_{0}^{+\infty} \frac{\d q}{q^{n-1}}
\frac{1}{q^4+\lambda^4}\int_0^{2\pi}(1-\cos^2\theta){\cal
U}_n(\cos\theta)\d\theta\;.
\end{eqnarray}
Since $n\geq 1$ and making use of \eqref{cheb}, for the
$\theta$-integration we find
\begin{eqnarray}
\int_0^{2\pi} (1-\cos^2\theta){\cal
U}_n(\cos\theta)\d\theta&=&\int_0^{2\pi}\sin\theta\sin((n+1)\theta)\d\theta\nonumber\\&=&\int_0^{2\pi}\frac{\cos(n\theta)-\cos((n+2)\theta)}{2}\d\theta=0\;.
\end{eqnarray}
However, this does not make the integral in \eqref{slecht} well
defined, as the remaining $q$-integral is infrared singular for any
occurring value of $n$! In fact, exactly these infrared divergences
forbid the interchange of integral and of the infinite sum. This is
a nice example of the fact that the integral of a infinite sum can
be well defined, whereas the (sum of the) individual integrals are
not.

When we first integrate exactly for any $k$ and then  expand in
powers of $k^2$, we do recover the meaningful result \eqref{zinvol}
at $k^2\sim0$.

\section{The gluon propagator and positivity violation}
Before turning to the conclusion, we would like to recall that
another typical feature of the Gribov-Zwanziger scenario is that the
gluon propagator vanishes at zero momentum. More precisely, ${\cal
D}(0)=0$. This implies a maximal violation of positivity, see e.g.
\cite{Dudal:2008sp}, thereby signalling that the gluon is an
unphysical degree of freedom and hence ``confined''.

In $3D$ and $4D$, we have shown that the effects originating from
the coupling of the operator
$\overline{\varphi}\varphi-\overline{\omega}\omega$ to the theory
gives a finite nonzero value to ${\cal D}(0)$, in accordance with
the lattice data \cite{Dudal:2008sp,Dudal:2008rm}. Notice, however,
that there is still a clear violation of positivity notwithstanding
that ${\cal D}(0)\neq0$. Our results were in qualitative agreement
with the available lattice data \cite{Dudal:2008sp,Dudal:2008rm}.

As we have argued already, we must discard
$\overline{\varphi}\varphi-\overline{\omega}\omega$ in $2D$.
Consequently, ${\cal D}(0)$ still vanishes in $2D$ at tree level due
to the Gribov mass, as it is immediately verified from
\begin{equation}\label{}
    {\cal D}(p^2)=\frac{p^2}{p^4+\lambda^4}\;.
\end{equation}
In principle, one could explicitly check whether this persists
beyond tree level order. However, this leads to quite complicated
loop calculations, as can be appreciated from the $4D$ or $3D$
counterpart done in \cite{Dudal:2008sp,Dudal:2008rm}, and therefore
we shall not pursue this here.

\section{Conclusion}
In this letter, we have discussed why it is not possible to
``refine'' the Gribov-Zwanziger action in $2D$, in contrast with the
$3D$ or $4D$ case. In the latter case, we have shown in recent work
\cite{Dudal:2007cw,Dudal:2008sp,Dudal:2008rm} that the inclusion of
dynamical effects related to a novel mass operator, constructed with
the additional field present in the Gribov-Zwanziger action, has a
profound influence on the infrared behaviour of the theory, and
considerably changes the usual Gribov-Zwanziger predictions. The
main conclusion is that the ghost propagator is not infrared
enhanced but retains its $\frac{1}{q^2}$ singularity in the deep
infrared, while the gluon propagator becomes finite and nonvanishing
at zero momentum. The usual Gribov-Zwanziger scenario predicts a
$1/k^4$ singularity for the ghost propagator, and a vanishing gluon
propagator at zero momentum, ${\cal D}(0)=0$. Surprisingly, lattice
data at large volumes are in compliance with the refined analytical
results presented in \cite{Dudal:2007cw,Dudal:2008sp,Dudal:2008rm}.
Since the lattice data in $2D$ still predicts an infrared enhanced
ghost and vanishing ${\cal D}(0)$
\cite{Cucchieri:2007rg,Cucchieri:2008fc,Maas:2007uv}, we were
motivated to discuss how this would fit into our refined
Gribov-Zwanziger scenario \cite{Dudal:2007cw,Dudal:2008sp}. We have
shown that it is not possible to couple the particular operator,
$\overline{\varphi}\varphi-\overline{\omega}\omega$, to the action
in $2D$, as it triggers serious infrared instabilities, which are
peculiar to the $2D$ case. Thence, the usual Gribov-Zwanziger
scenario is so to say ``protected'' in $2D$. In fact, we have proven
that the emerging infrared singularities make it impossible to stay
within the Gribov region $\Omega$ when $M^2\neq0$. As a nice
byproduct of this work, we have seen that the Gribov mass can act as
a natural infrared regulator, stabilizing the otherwise ill-defined
perturbative expansion.

\section*{Acknowledgments}
We are grateful to A.~Maas and J.~M.~Pawlowski who motivated us to
investigate the $2D$ case. The Conselho Nacional de Desenvolvimento
Cient\'{i}fico e Tecnol\'{o}gico (CNPq-Brazil), the Faperj,
Funda{\c{c}}{\~{a}}o de Amparo {\`{a}} Pesquisa do Estado do Rio de
Janeiro, the SR2-UERJ and the Coordena{\c{c}}{\~{a}}o de
Aperfei{\c{c}}oamento de Pessoal de N{\'{i}}vel Superior (CAPES) are
gratefully acknowledged for financial support. D.~Dudal and
N.~Vandersickel acknowledge the financial support from the Research
Foundation - Flanders (FWO). N.~Vandersickel is grateful for the
hospitality at the CTP, MIT where this work was completed. This work
is supported in part by funds provided by the US Department of
Energy (DOE) under cooperative research agreement DEFG02-05ER41360.

\end{document}